\documentstyle[12pt]{article}
\date{}
\begin{document}

\title{Representations of General Dimension for the Skyrme Model}

\author{E. Norvai\v{s}as$^1$ and D.O. Riska$^2$}
\maketitle

\centerline{\it $^1$Institute of Theoretical Physics and Astronomy,
Vilnius, 2600 Lithuania}

\centerline{\it $^2$Department of Physics, University of Helsinki,
00014 Finland}

\setcounter{page} {0}
\vspace{1cm}

\centerline{\bf Abstract}

We construct the representations of general dimension for the soliton
solution to the $SU(2)$ Skyrme model, and show that at the classical
level the dependence on the dimension of the
representation ($2j+1$) appears only as an overall factor $j(j+1)(2j+1)$
in the Lagrangian density, which may be absorbed by a rescaling of the
parameters. Alternate stabilizing terms in the model will in general have a
different $j$-dependence and have to be rescaled accordingly in order
to achieve representation independent predictions. In contrast the
quantum corrections do depend on the
dimension of the representation and in general differ from those obtained
in the fundamental representation of $SU(2)$.

\newpage

\centerline{\bf 1. Introduction}
\vspace{0.5cm}

Skyrme's topological soliton model for the baryons [1,2] and its
immediate generalizations [3] have proven able to provide a
qualitatively remarkably succesful description of most of the observed
properties of the nucleons, including the hyperons [4,5]. The model is
formed by a Lagrangian density for an $SU(2)$ field $U$, and
topological baryon current $B^\mu$, which is conserved by the unitarity
condition $UU^\dagger=1$ independently of the form of the
Lagrangian.\\

The soliton solution is obtained by the hedgehog ansatz
$$
U_0=e^{i\vec \tau\cdot \hat r F(r)},\eqno(1.1)
$$
where $\vec \tau$ is a Pauli-isospin matrix and $F(r)$ a scalar
function, which satisfies a second order differential equation that is
obtained by the requirement that the solution lead to a stationary
energy. The spherical components of the operator ${1\over 2} \vec
\tau$ form the generators of the fundamental representation of the
group $SU(2)$, and satisfy the commutation relations
$$
[\hat J_a,\, \hat J_b]=\left[ \begin{array}{ccc}1 & 1 & 1\\a & b &
c\end{array} \right] \hat J_c,\eqno(1.2)
$$
where the factor on the r.h.s. is the Clebsch-Gordan coefficient
$(1a1b|1c)$, in a more convenient notation. Here we have used the
normalization $\hat J_{\pm}=-J_{\pm 1}/\sqrt{2},\, \hat J_0=-J_0/\sqrt{2}$.\\

In this paper we shall form the analogs to the hedgehog solution (1.2)
for representations of arbitrary integral dimension and derive the
corresponding expressions for the soliton mass. At the classical level
we find that the predictions for the static baryon observables can be made
independent of the dimension of the representation by a dimension
dependent rescaling of the parameters of the model. In the case of the
original Skyrme model the dependence on the dimension $(2j+1)$ of the
representation appears only in the form of an overall factor $j(j+1)(2j+1)$.
The quantized
Hamiltonian for the soliton, will in contrast depend
nontrivially on the
dimensionality. This opens up a hitherto unexplored phenomenological
degree of freedom, which may be of dynamical significance.\\

In section 2 of this note we express the Skyrme model and its most direct
generalizations in a general
representation of the group $SU(2)$ and derive the explicit
representation dependence of the terms in the Lagrangian density at
the classical level.
In section 3 we derive the expression for the hedgehog
solution and the corresponding soliton mass for the case of a
representation of arbitrary order. In section 4 we consider the
quantized version of the model. Section 5 contains a summarizing
discussion.\\

\vspace{1cm}

{\bf 2. The Skyrme model in a general representation}\\
\vspace{0.5cm}

In an irreducible representation of the group $SU(2)$ of dimension
$j$ the unitary field
$U$ has the form
$$
U(\vec r,t)=D^j(\vec \alpha (\vec r,t)),\eqno(2.1)
$$
where $\vec \alpha (\vec r,t)$ is a vector formed of three Euler angles
$\vec \alpha\equiv (\alpha^1,\, \alpha^2,\, \alpha^3)$:
$$
0\leq \alpha^1 < 2\pi,\quad 0\leq \alpha^2 \leq \pi,\quad 0\leq
\alpha^3 <4\pi,\eqno(2.2)
$$
and $D^j$ is the Wigner $D$-function, that has the explicit form
$$
D_{mm'}^{j}(\vec \alpha)=\langle jm|e^{i\sqrt{2}\,\alpha^1
\hat J_0}\,e^{-\alpha^2(\hat J_{+} +\hat J_{-})}\,e^{i\sqrt{2}\,\alpha^3 \hat
J_0}
|jm'\rangle .\eqno(2.3)
$$
The Euler angles $\vec \alpha$ then form the dynamical variables of
the theory. Note that the trace of a bilinear combination of two
generators of the groups depends on the dimension of the representation
$(j)$ as
$$
Tr\langle jm|J_a J_b|jm'\rangle =(-)^a{1\over
6}j(j+1)(2j+1)\delta_{a,-b}.\eqno(2.4)
$$
The Skyrme model is the chirally symmetric Lagrangian density
$$
{\cal L}=-{f_\pi^2 \over 4} Tr\{R_\mu R^\mu\} +{1\over 32
e^2}Tr\{[R_\mu,R_\nu]^2\},\eqno(2.5)
$$
where the "right" current $R_\mu$ is defined as
$$
R_\mu=(\partial_\mu U)U^\dagger.\eqno(2.6)
$$
and $f_\pi$ (the pion decay constant) and $e$ are parameters. For the
purpose of expressing the Lagrangian density in terms of the Euler
angles $\{\vec \alpha\}$ it is convenient to note that
$$
{\partial \over \partial \alpha^i}\,D_{mn}^{j}(\vec
\alpha)=C_i^{(a)}(\vec \alpha)\,\langle jm|\hat J_a|jm'\rangle
\,D_{m'n}^{j}(\vec
\alpha).\eqno(2.7)
$$
Here the coefficients $C_i^{(a)}(\vec \alpha)$ have the explicit form
$$
\begin{array}{lll}
\quad C_1^{(+)}(\vec \alpha)=0,&\quad C_2^{(+)}(\vec \alpha)=-e^{-i\alpha^1},%
&\quad C_3^{(+)}(\vec \alpha)=-i \sin \alpha^2\, e^{-i\alpha^1},\\
\quad C_1^{(0)}(\vec \alpha)=i\sqrt{2},&\quad C_2^{(0)}(\vec \alpha)=0,&%
\quad C_3^{(0)}(\vec \alpha)=i\sqrt{2}\cos \alpha^2,\\
\quad C_1^{(-)}(\vec \alpha)=0,&\quad C_2^{(-)}(\vec \alpha)=-e^{i\alpha^1},&%
\quad C_3^{(-)}(\vec \alpha)=i \sin \alpha^2\, e^{i\alpha^1}.
\end{array}\eqno(2.8)
$$
The right current $R_\mu$ then takes the form
$$
(R_\mu)_{mm'}=\partial_\mu \alpha^i C_i^{(a)}(\vec
\alpha)\,\langle jm|\hat J_a|jm'\rangle ,\eqno(2.9)
$$
where summation over the indices $i$ and $a$ is understood.

The chiral invariance of the theory is the invariance under the
transformation
$$
U(\vec r,t)\rightarrow VU(\vec r,t)W^\dagger,\eqno(2.10)
$$
in the $(j,j)$ representation of $SU(2)\times SU(2)$, where the
left and right transformation matrices $V$ and $W^\dagger$ belong to
the irreducible representations of the product groups. Under the left
chiral transformation the right current $R_\mu$ transforms as
$$
R'_\mu=VR_\mu\, V^\dagger=\partial_\mu \alpha^i C_i^{(a)} (\vec
\alpha)\,\langle j|\hat J_{a'}|j\rangle \,D_{a'a}^{1}(\vec \beta).\eqno(2.11)
$$
Here $\vec \beta$ are the Euler angles that define the transformation
matrix $V$.\\

When reexpressed in terms of the Euler angles $\vec \alpha$ the
Lagrangian density (2.4) takes the form
$$
{\cal L}={1\over 3} j(j+1)(2j+1)\bigg\lbrace {f_\pi^2 \over 4}\Bigl[
\partial_\mu
\alpha^i\partial^\mu \alpha^i
+2\cos \alpha^2\, \partial_\mu \alpha^1 \, \partial^\mu \alpha^3\Bigr]
$$
$$
-{1\over 16e^2}\Bigl[ \partial_\mu \alpha^2\partial^\mu
\alpha^2(\partial_\nu \alpha^1 \partial^\nu \alpha^1+\partial_\nu
\alpha^3\partial^\nu \alpha^3)
-(\partial_\mu \alpha^1 \partial^\mu \alpha^2)^2\bigskip
$$
$$
-(\partial_\mu \alpha^2 \partial^\mu \alpha^3)^2
+\sin^2\!\alpha^2[\partial_\mu \alpha^1 \partial^\mu \alpha^1 \,
\partial_\nu \alpha^3 \partial^\nu \alpha^3-(\partial_\mu \alpha^1
\partial^\mu \alpha^3)^2]\bigskip
$$
$$
+2\cos \alpha^2[\partial_\mu \alpha^2 \partial^\mu \alpha^2\,
\partial_\nu \alpha^1 \partial^\nu \alpha^3-\partial_\mu \alpha^1
\partial^\mu \alpha^2 \partial_\nu \alpha^2\partial^\nu
\alpha^3] \Bigr] \bigg\rbrace .\eqno(2.12)
$$
Note that the only dependence on the dimension of the representation
is in the overall factor $j(j+1)(2j+1)$ in the Lagrangian density.
This implies that the equation of motion for the dynamical variable
$\vec \alpha$ is independent of the dimension of the representation as
the common factor can be absorbed into the two parameters of the model.\\

The conserved topological current in the Skyrme model is the baryon
current
$$
B^\mu=N \epsilon^{\mu \nu \beta \gamma} \,Tr\, R_\nu \, R_\beta\,
R_\gamma,\eqno(2.13)
$$
where the normalization factor $N$ depends on the dimension of the
representation and has the value $1/24\pi^2$ in the fundamental
representation. The baryon number $B$ is obtained as the spatial
integral of the time component $B^0$. In terms of the Euler angle
variables $\vec\alpha$ the baryon current takes the form
$$
B^\mu=-{N\over 6}j(j+1)(2j+1)\sin \alpha^2 \,\epsilon^{\mu \nu \beta
\gamma}\, \epsilon_{ikl}\,\partial_\nu \alpha^i\partial_\beta
\alpha^k\partial_\gamma \alpha^l.\eqno(2.14)
$$
As the dimensionality of the representation appears in this expression
in the same overall factor as in the Lagrangian density (2.11) it
follows that all calculated dynamical observables will be independent
of the dimension of the representation at the classical level, if this
factor $j(j+1)(2j+1)$ is absorbed into the parameters of the model.\\

There exists an infinite class of alternate stabilizing terms for the
Lagrangian density (2.5), combinations of which can be used in place
of Skyrme's quartic stabilizing term or be added to it [3]. An
alternate term of quartic order, which leads to identical results as
the Skyrme term in the fundamental representation, is [6]
$$
{\cal L}'_4={1\over 16 e^2}\{(Tr\, R_\mu\, R_\nu)^2-(Tr\, R_\mu\,
R^\mu)^2\}.\eqno(2.15)
$$
When this term is expressed in terms of the Euler angles $\{\vec
\alpha\}$ (2.1), the resulting Lagrangian density has the form (2.12),
with the exception that stabilizing term that is proportional to $ $
$e^{-2}$ $ $ has an additional factor $ $ ${2\over 3}$ $j(j+1)(2j+1)$. Hence
invariance of the physical predictions requires that the parameter
$(1/e^2)$ of the stabilizing term (2.15) be taken to be proportional to
$[j(j+1)(2j+1)]^{-2}$, and the parameter $(f_\pi)$ of the quadratic
term to be proportional to $[j(j+1)(2j+1)]^{-1}$, when a
representation of dimension $j$ is employed.

Consider then the sixth order stabilizing term [3,7]
$$
{\cal L}_6=e_6\, Tr\{[R_\mu,\, R^\nu][R_\nu,\,
R^\lambda][R_\lambda,\, R_\mu]\}.\eqno(2.16)
$$
\newpage
In terms of the Euler angles $\{ \vec \alpha\}$ this Lagrangian
density takes the form
$$
{\cal L}_6=-\frac{j(j+1)(2j+1)}{6}e_6\,\epsilon_{i_1i_2i_5}\,
\epsilon_{i_3i_4i_6}\, \sin^2\!\alpha^2
$$
$$
\partial_\mu
\alpha^{i_1}\, \partial^\nu\alpha^{i_2}\, \partial_\nu\alpha^{i_3}
\partial^\lambda \alpha^{i_4}\, \partial_\lambda\alpha^{i_5}\,
\partial^\mu \alpha^{i_6}\,
.\eqno(2.17)
$$
This result reveals that the dependence on the dimension of the
representation of this term is contained in the same overall factor
$j(j+1)(2j+1)$ as the Skyrme model Lagrangian (2.12). Hence addition
of the term ${\cal L}_6$ maintains the simple overall dimension
dependent factor of the original Skyrme model.\\

As in the case of the quartic term one can construct an alternate
sixth order term, which is equivalent to (2.16) in the case of the
fundamental representation, but which differs in its dependence on the
dimension $j$:
$$
{\cal L}'_6=e'_6\, \epsilon^{\mu
\nu_1\nu_2\nu_3}\,\epsilon_{\mu\eta_1\eta_2\eta_3}\,
Tr\{R_{\nu_1}R_{\nu_2}R_{\nu_3}R^{\eta_1}R^{\eta_2}
R^{\eta_3}\}.\eqno(2.18)
$$
In terms of the Euler angles $\{\vec \alpha\}$ this term also reduces
to the expression (2.17), with the exception of an additional factor
$j(j+1)(2j+1)e'_6/e_6$. Its dependence on $j$ is thus different from (2.16),
although by adjusting the values of the parameters $e_6$ and $e'_6$
differently in each representation equivalent dynamical predictions
can be maintained.

\vspace{1cm}

{\bf 3. The hedgehog solution in a general representation}\\
\vspace{0.5cm}

The hedgehog field (1.1) represents the soliton solution in the
fundamental representation of $SU(2)$. In order to find its
generalizations in representations of higher dimension one may compare
it to the matrix elements $D_{mm'}^{1/2}(\vec \alpha)$, and thus obtain
the explicit
expressions for the Euler angles $\vec \alpha$ in terms of the chiral
angle $F(r)$. The result is
$$
\begin{array}{l}
\alpha^1=\varphi-\arctan (\cos \vartheta \, \tan F(r))-\pi/2,\\
\alpha^2=-2\arcsin(\sin \vartheta\, \sin F(r)), \\
\alpha^3=-\varphi-\arctan(\cos \vartheta\, \tan F)+\pi/2.
\end{array}\eqno(3.1)
$$
Here the angles $\varphi, \vartheta$ are the polar angles that define the
direction of the unit vector $\hat r$ in spherical coordinates.\\

Substitution of the expressions (3.1) into the general expression
(2.1) for the unitary field $U$ then gives the hedgehog field in a
representation with arbitrary $j$. As an example the hedgehog field in
the representation $j=1$ has the form
$$
U_0=\sin^2\!F \left( \begin{array}{ccc} G^2 & i\sqrt{2} G \sin
\vartheta\, e^{-i\varphi} & -\sin^2 \!\vartheta\, e^{-2i \varphi}\\
i\sqrt{2} G \sin \vartheta\, e^{i\varphi} & \cot^2 F+\cos 2 \vartheta &
i\sqrt{2}\sin \vartheta G^*\, e^{-i\varphi}\\-\sin^2 \!\vartheta
e^{2i\varphi} & i\sqrt{2} G^* \sin \vartheta\, e^{i\varphi} & G^{*2}
\end{array} \right) .\eqno(3.2)
$$
Here we have used the abbrevations
$$
G=\cot \, F+i\cos \vartheta.\eqno(3.3)
$$
The Lagrangian density (2.12) reduces to the following simple form,
when the hedgehog ansatz (3.1) is employed:
$$
{\cal L}=-{4\over 3}j(j+1)(2j+1)\bigg\{{f_\pi^2\over 4}\Bigl( F^{'2}+{2\over
r^2}\sin^2 \!F\Bigr)
$$
$$
+{1\over 16e^2}{\sin^2 \!F \over r^2}\Bigl( 2F^{'2}+{\sin^2 \!F\over
r^2}\Bigr) \bigg\}.\eqno(3.4)
$$
For $j=1/2$ this reduces to the result of ref. [2]. The corresponding
mass density is obtained by reversing the sign of ${\cal L}$.\\

The requirement that the soliton mass be stationary yields the
equation of motion for the chiral angle $F$ [2,8]:
$$
f_\pi^2\Bigl( F''+{2\over r}F'-{\sin 2F\over r^2}\Bigr) -{1\over e^2}
\Bigl( {1\over
r^4}\sin^2\!F\,\sin 2F
$$
$$
-{1\over r^2} (F^{'2}\sin 2F+2F'' \sin^2\!F)\Bigr) =0,\eqno(3.5)
$$
which is independent of the dimension of the representation. Note that
this differential equation is nonsingular only if $F(0)$ is an integer
multiple of $\pi$. \\

For the hedgehog form the baryon density reduces to the expression
$$
B^0=-8Nj(j+1)(2j+1)\frac{\sin^2 F}{r^2} F'.\eqno(3.6)
$$
The corresponding baryon number is
$$
B=\int d^3 rB^0=16N\pi j(j+1)(2j+1)[F(0)-{1\over 2}\sin
2F(0)].\eqno(3.7)
$$
Combining the requirement that $F(0)$ be an integer multiple of $\pi$
with the requirement that the lowest nonvanishing baryon number be 1
gives the general expression for the normalization factor $N$ as
$$
N={1\over 16\pi^2 j(j+1)(2j+1)},\eqno(3.8)
$$
which reduces to the usual result $1/24 \pi^2$ for $j=1/2$.

\vspace{1cm}

{\bf 4. The quantized Lagrangian density}
\vspace{0.5cm}

In the quantized version of the Lagrangian density (2.5) of the Skyrme
model (and the generalizations of it considered in section 2 above)
the analogs of the Euler angles $\{\vec \alpha\}$ that define the
$SU(2)$ matrices $U$ are a set of three real parameters $\{\vec
q\}$, which satisfy the general commutation relations [9]
$$
[\dot{q}^a,\, q^b]=-if^{ab}(\vec q).\eqno(4.1)
$$
Here the tensor $f^{ab}$ is a function of the generalized coordinates
$\{\vec
q\}$ only, the explicit form of
which is determined after the quantization condition has
been imposed. Explicit expressions
for $f^{ab}$ have given in ref.[9] for the case of the fundamental
representation using the method of collective coordinates in
the quantization. The tensor $f^{ab}$ is symmetric with respect to
interchange of the indices $a$ and $b$ as a consequence of the
commutation relation $[q^a,\, q^b]=0$.  The commutator between a
generalized veloicity component $\dot{q}^a$ and a function $F$ of the
coordinates is given by
$$
[\dot{q}_a,\, F(\vec q)]=-i\sum_{r}f^{ar}(\vec q){\partial \over
\partial q^r}F(\vec q)=-i\sum_{r}f^{ar}(\vec q)\,\nabla _{\!\!r}\,F(\vec
q).\eqno(4.2)
$$
The commutation relation (4.1) leads to complicated expressions for
the time derivatives of the $SU(2)$ matrices $U$, $U^\dagger$, which
appear in the time components of the right invariant current $R$
(2.6). In a representation of the group $SU(2)$ of dimension $j$ we
find the expression
$$
{(R_0)}^j_{mn}={1\over 2}\{\dot{q}^i,\, {C}_i^{(a)}(\vec q)\}\,\langle jm|\hat
J_a|jn\rangle \medskip
$$
$$
+{1\over 2} if^{kd}(\vec q)\,C_k^{(a)}(\vec q)\,C_d^{(b)}(\vec
q)\sum_{l=0,2}\left[\begin{array}{ccc} 1 & 1 & l\\ a & b &
u\end{array}\right]\langle jm|\hat J_u^l|jn\rangle .\eqno(4.3)
$$
Here $\{,\}$ denotes an anticommutator. Whereas in the classical case
the currents form an $SU(2)$ algebra, the time components of the
currents in the quantum mechanical case
belong to the product space of two generators, the elements
of which can be decomposed into sums of tensors of rank $0,1$ and $2$
as indicated in the r.h.s. of (4.3).

The expression (4.3) implies that the fundamental representation
represents a special case, because
in it the generators can be expressed as Pauli
matrices, the algebra of which is exceptionally simple, and for which
$$
\langle {1\over 2}m|\hat J_u^2|{1\over 2} n\rangle =0,\eqno(4.4)
$$
so that the rank two tensors absent in this representation.\\

The quantal analog of the chiral transformation law (2.11) is
$$
(R'_0)^j_{mn}={1\over 2}\{\dot{q}^2,\, C_i^{(a)}(\vec q)\}\langle jm|\hat
J_{a'}|jn\rangle D_{a'a}^{1}(\vec \beta)
$$
$$
+{i\over 2}f^{kd}(\vec q)\,C_k^{(a)}(\vec q)\,C_d^{(b)}(\vec
q)\sum_{l=0,2}\left[\begin{array}{ccc} 1 & 1 & l\\ a & b & u
\end{array}\right]\,\langle jm|\hat J_{u'}^{l}|jn\rangle \,D_{u'u}^{l}(\vec
\beta),\eqno(4.5)
$$
where $\vec \beta$ are the Euler angles that define the transformation
matrix.\\

In the quantum mechanical case the Lagrangian density (2.5) (as well
as the additional terms (2.15)-(2.17)) remain invariant under chiral
transformations (2.10), (4.5). The explicit expressions will in this
case depend on the tensor $f^{ab}(\vec q)$ that determines the
commutation relation (4.1). The quantum mechanical expression for the
quadratic term which depends on time derivatives in the Lagrangian density is
$$
{\cal L}_{2}^t=-{f_\pi^2\over 24}j(j+1)(2j+1)\Big\{\dot{q}^a\, g_{ab}\,
\dot{q}^b
$$
$$
-{1\over 4}f^{aa'}\nabla_{\!\!a'}(f^{bb'}\nabla_{\!\!b'}g_{ab})
-\frac{j(j+1)}{24(2j+1)}f^{ab}g_{ab}f^{a'b'}\,g_{a'b'}
$$
$$-{1\over 80}(2j-1)(2j+3)f^{ab}C_{a}^{(k)}C_{b}^{(l)}\left[
\begin{array}{ccc} 1 & 1 & 2\\k & l & k+l\end{array}\right]
$$
$$
\times f^{a'b'}C_{a'}^{(m)}C_{b'}^{(-k-l-m)}\left[ \begin{array}{ccc} 1 & 1
& 2\\ m & -k-l-m & -k-l\end{array}\right]\Big\}.\eqno(4.6)
$$
Here we have definded the $3\times 3$ tensor $g_{ab}$ as the scalar
product
$$
g_{ab}=\sum_{m}(-)^m C_{a}^{(m)} C_{b}^{(-m)}=-2\delta _{ab}
-2(\delta _{a1} \delta _{b3} +\delta_{a3} \delta_{b1} )\cos q^2.\eqno(4.7)
$$
In terms of the quantum mechanical variables $\{\vec q\}$ the quartic
term of the Lagrangian density (2.5) is
$$
{\cal L}_{4}^t={1\over
192e^2}j(j+1)(2j+1)\Bigg\{ \dot{q}^a\tilde{g}_{ab}\,\dot{q}^b
-{1\over 4}f^{bb'}\nabla_{\!\!b'}(f^{aa'}\nabla_{\!\!a'}\tilde{g}_{ab})
$$
$$
-{1\over 6}{j(j+1)\over (2j+1)}(-)^{k+m}\partial_\mu
q^{a'}\partial^\mu q^{b'}C_{a}^{(k)} C_{b}^{(m)}f^{aa''}\nabla_{\!\!a''}
C_{a'}^{(-k)}f^{bb''}\nabla_{\!\!b''}C_{b'}^{(-m)}
$$
$$
+{(-)^{1+k}\over 20} (2j-1)(2j+3)\partial_\mu q^{a'}\partial^\mu
q^{b'}
$$
$$
\times \left( \sqrt{6}C_{a'}^{(k')}f^{aa''}C_{a}^{(k-k'-k'')}C_{a''}^{(k'')}
 \left[\begin{array}{ccc}1
 & 1 & 2\\k-k'-k'' & k'' & k-k'\end{array}\right]
\left[\begin{array}{ccc}2 & 1 & 2\\k-k' & k' & k\end{array}\right]\right.
$$
$$
-2C_{a}^{(k-k')}f^{aa''}\nabla_{\!\!a''}C_{a'}^{(k')}\left[\begin{array}{ccc}1
 & 1 & 2\\k-k' & k' & k\end{array} \right]\biggr)
$$
$$
\times
\left(\sqrt{6}C_{b'}^{(m')}f^{bb''}C_{b}^{(-k-m'-m'')}C_{b''}^{(m'')}\left[
\begin{array}{ccc}1 & 1 & 2\\-k-m'-m'' & m'' &
-k-m'\end{array}\right]\right.
$$
$$
\times \left.\left[\begin{array}{ccc}2 & 1 & 2\\k-m' & m' &
-k\end{array}\right]
-2C_{b}^{(-k-m')}f^{bb''}\nabla_{\!\!b''}C_{b'}^{(m')}\left[\begin{array}{ccc}1
 & 1 & 2\\-k-m' & m' & -k\end{array}\right]\right)\Bigg\}.\eqno(4.8)
$$
Here the tensor $\tilde{g}_{ab}$ has been defined as
$$
\tilde{g}_{ab}=(-)^k\partial_\mu q^{a'}\partial^\mu
q^{b'}(\nabla_{\!\!a}C_{a'}^{(k)}-\nabla_{\!\!a'}C_{a}^{(k)})
(\nabla_{\!\!b}C_{b'}^{(-k)}-\nabla_{\!\!b'}C_{b}^{(-k)})
$$
$$
=(-)^k \partial_\mu q^{a'} \partial^\mu q^{b'}
C_{a}^{(k')}C^{(k-k')}_{a'}C_{b}^{(k'')}C_{b'}^{(-k-k'')}\medskip
$$
$$
\times \left[\begin{array}{ccc}1
 & 1 & 1\\k' & k-k' & k\end{array}\right]
\left[\begin{array}{ccc}1 & 1 & 1\\k'' &-k-k'' &-
k\end{array}\right].\eqno(4.9)
$$

The terms without time derivatives are the same as in classical case (2.12).
In the fundamental representation ($j=1/2$) the last two terms in the
Lagrangian densities (4.6) and (4.8), which contain the factor
$(2j-1)$ vanish. In this case using the method of collective coordinates
the quantized Lagrangian of the Skyrme
model reduces to the case considered in ref. [9]. It is the presence
of those terms which leads to the essential inequivalence of the
quantized Skyrme model Lagrangians that appear in representations of
higher dimensionality and that in the fundamental representation.

In the case of higher symmetry groups than $SU(2)$
the Wess-Zumino
action
$$
S=i{N_c\over 240\pi^2} \int d^5x\, \epsilon^{\mu \nu \alpha \beta
\gamma}\,Tr\, R_\mu R_\nu R_\alpha R_\beta R_\gamma\eqno(4.10)
$$
(here written with the normalization appropriate for the fundamental
representation [2]) also contributes to the energy. This
has to be added to the Skyrme model
Lagrangian in order to break its discrete reflection symmetry, which is
not a symmetry of $QCD$.
While the proof that it vanishes in a general representation of $SU(2)$
is somewhat involved it
is straightforward
in the case of the fundamental representation
for $SU(2)$ because all combinations of
its generators (Pauli matrices) can be reduced to linear combinations
of the three generators and a scalar, and therefore at least two of
the 5 current operators $R$ will have equal space-time indices, and
consequently the expression (4.10) vanishes by antisymmetry.
In the quantum case one of current operators $R$ can be chosen in the form
(4.3). Also the expression (4.10) vanishes by antisymmetry of three Euler
angles.
\vspace {1cm}

{\bf 5. Discussion}
\vspace {0.5 cm}

All phenonomenological applications of the Skyrme model and its
extensions to the description of the structure of baryons and nuclei
have hitherto relied on the fundamental representation of $SU(2)$.
Above we have shown that substitution of a higher dimensional
irreducible representation in place of the fundamental one does not
change the phenomenological predictions obtained with the classical
version of the model provided the parameters of the Lagrangian
density are rescaled in an appropriate dimension dependent way.

In the quantum mechanical version of the model the representation
dependence cannot however be eliminated by a simple rescaling
of the parameters.
The difference between the
fundamental representation and those of higher dimension is
the appearance of tensors of rank 2 in the transformation
law for the currents $U\partial^\mu U^\dagger$. Thus the
baryon spectra that are obtained in the fundamental representation
with the collective coordinate quantization method depend in
an essential way on the exceptionally simple algebraic
properties of the generators of that representation.

We have derived explicit expressions for the Lagrangian density
and the currents of the Skyrme model for representations of
arbitrary integral dimension $2j+1$. The derivation of the
baryon spectra in the case of an arbitrary representation
in addition requires a choice of quantization method. It shall
be an interesting question to determine the particle spectra
in the case of the higher dimensional representation using the
quantization method of collective variables.

{\bf Acknowledgements}

E.N. acknowledges partial support by the International Science
Foundation and travel support by NORDITA.
\newpage

{\bf References}\\
\vspace{0.5cm}

\begin{enumerate}
\item T.H.R. Skyrme, Proc. Roy. Soc. {\bf A260} (1961) 127
\item G.S. Adkins, C.R. Nappi and E. Witten, Nucl. Phys. {\bf B228}
(1983) 552
\item L. Marleau, Phys. Rev. {\bf D45} (1992) 1776
\item C.G. Callan, K. Hornbostel and I. Klebanov, Phys. Lett. {\bf
B202} (1988) 269
\item M. Rho, D.O. Riska and N.N. Scoccola, Z. Phys. {\bf A341} (1992)
343
\item G. Pari, Phys. Lett. {\bf B261} (1991) 347
\item A. Jackson et al., Phys. Lett. {\bf 154B} (1985) 101
\item E.M. Nyman and D.O. Riska, Rept. Prog. Phys. {\bf 53} (1990)
1137
\item K. Fujii, A. Kobushkin, K. Sato and N. Toyota, Phys. Rev. {\bf
D35} (1987) 1896
\end{enumerate}
\end{document}